\documentclass[12pt]{article}
\usepackage{amsmath, mathtools}
\usepackage{graphicx}
\usepackage{enumerate}
\usepackage{natbib}
\usepackage{url} 
\usepackage{amsfonts}
\usepackage{amsmath}
\usepackage{amssymb}
\usepackage{amsthm}
\usepackage{enumerate}
\usepackage{graphicx}
\usepackage{layout}
\usepackage{latexsym}
\usepackage{array}
\usepackage{extarrows}
\usepackage{balance}
\usepackage{titlesec}
\usepackage{bbm}
\usepackage{floatrow}  
\usepackage{rotating} 
\usepackage{color}
\usepackage{dcolumn}
\usepackage{multicol}
\usepackage{multicol}
\usepackage{setspace}
\usepackage{listings}
\usepackage{fancyhdr}
\usepackage{float}
\usepackage{graphicx}
\usepackage{subfigure}
\floatstyle{plaintop}
\restylefloat{table}
\usepackage[flushleft]{threeparttable}
\usepackage[CJKbookmarks=true]{hyperref}
\hypersetup{hidelinks}
\usepackage{hyperref}

\newtheorem{theorem}{Theorem}

\newtheorem{remark}{Remark}

\newtheoremstyle{procedure}{3pt}{3pt}{\upshape}{}{}{}{}{}

\def\argmin{\mathop{\rm argmin}\limits}

\newcommand{\bcon}{{\bf Condition }}

\newcommand{\bdelta}{\boldsymbol{\delta}}
\newcommand{\bbeta}{\boldsymbol{\beta}}
\newcommand{\hatbbeta}{\widehat{\boldsymbol{\beta}}}

\newcommand{\btheta}{\boldsymbol{\theta}}

\newcommand{\bX}{\mathbf{X}}

\newcommand{\bx}{\mathbf{x}}
\newcommand{\bZ}{\mathbf{Z}}
\newcommand{\bz}{\mathbf{z}}

\newcommand{\blind}{1}

\usepackage{xcolor}
\def\boxit#1{\vbox{\hrule\hbox{\vrule\kern1pt
			\vbox{\kern1pt#1\kern1pt}\kern1pt\vrule}\hrule}}

\definecolor{dartmouthgreen}{rgb}{0.05, 0.5, 0.06}

\begin{document}

	\def\spacingset#1{\renewcommand{\baselinestretch}%
		{#1}\small\normalsize} \spacingset{1}

	
	\if1\blind
	{
		\title{\bf High-Dimensional Extreme Quantile Regression}
		\author{Yiwei Tang\\
			Department of Statistics and Data Science, Fudan University\\
			and \\
			Huixia Judy Wang \\
			Department of Statistics, The George Washington University\\
			and \\
			Deyuan Li \\
			Department of Statistics and Data Science, Fudan University}
		\maketitle
	} \fi
	
	\if0\blind
	{
		\bigskip
		\bigskip
		\bigskip
		\begin{center}
			{\LARGE\bf High-Dimensional Extreme Quantile Regression}
		\end{center}
		\medskip
	} \fi
	
	\bigskip
	\begin{abstract}
		The estimation of conditional quantiles at extreme tails is of great interest in numerous applications. Various methods that integrate regression analysis with an extrapolation strategy derived from extreme value theory have been proposed to estimate extreme conditional quantiles in scenarios with a fixed number of covariates. However, these methods prove ineffective in high-dimensional settings, where the number of covariates increases with the sample size. In this article, we develop new estimation methods tailored for extreme conditional quantiles with high-dimensional covariates. We establish the asymptotic properties of the proposed estimators and demonstrate their superior performance through simulation studies, particularly in scenarios of growing dimension and high dimension where existing methods may fail. Furthermore, the analysis of auto insurance data validates the efficacy of our methods in estimating extreme conditional insurance claims and selecting important variables.
	\end{abstract}
	
	\noindent%
	\vfill
	
	\newpage
	\spacingset{1.9} 
	\section{Introduction}
	\label{sec:intro}
	
	Quantile regression has become a widely recognized and useful alternative to classical least-squares regression for analyzing heterogeneous data. Since its introduction by \cite{koenker1978regression}, quantile regression has gradually been extended to a wide variety of data analytic settings; for a comprehensive review, see \cite{koenker2005quantile, koenker2017handbook}.
	
	While traditional quantile regression allows exploration of a wide range of conditional quantiles for $ \tau\in[\tau_l,\tau_u]$ with $\tau_l,\tau_u\in(0,1)$, there is often interest in the extreme tails where $\tau$ is close to 0 or 1. Without loss of generality, we focus the discussion on $\tau$ close to 1. The inherent challenge in estimating tail quantiles lies in the fact that the number of observations in the tails, that is, above the 
	$\tau$th quantile, is often small. In various contexts, methods have been developed to study extreme quantile regression by leveraging extreme value theory. \cite{chernozhukov2005extremal}, \cite{wang2012estimation}, and \cite{wang2013estimation}  studied extreme conditional quantiles under the linear quantile regression framework, and \cite{li2019extreme} focused on the linear quantile autoregressive model. \cite{daouia2013kernel} and \cite{gardes2019integrated} considered nonparametric smoothing methods. \cite{wang2009tail} and \cite{youngman2019generalized} applied generalized Pareto models to analyze the exceedance at a high threshold.  \cite{velthoen2023gradient} and \cite{gnecco2024extremal} explored tree-based methods such as gradient boosting and random forest.

	With advancements in data collection techniques such as genomics, economics, finance, and imaging studies,  the dimension of covariates $p$ is becoming larger and can grow with the sample size $n$. Existing methodology and theory in quantile regression for high-dimensional covariates have primarily focused on a central quantile level or compact quantile sets in $(0,1)$. {In situations where $p$ grows with $n$, various studies have investigated asymptotic behaviors of quantile regression estimators, e.g., \cite{welsh1989m}, \cite{he2000parameters}, \cite{belloni2019conditional}, \cite{pan2021multiplier}, and \cite{he2023smoothed}.  Assuming the sparsity in regression coefficients, researchers have explored the $\ell_1$-penalized quantile regression estimator and its generalizations \citep{belloni2011ℓ1, fan2014adaptive}, concave penalties \citep{wang2012quantile}, and smoothed quantile regression with concave penalties \citep{tan2022high}. Additionally, the debiased estimator of high-dimensional quantile regression has been studied for inference at central quantiles \citep{zhao2014general, bradic2017uniform}.}
	
	
	In this article, we address the challenge of estimating extreme conditional quantiles of $Y\in \mathbb{R}$ given a set of predictors $\bX\in\mathbb{R}^p$ in a high-dimensional context. The challenge lies in the scarcity of efficient samples at the tail and the sparsity of samples in high-dimensional space. As noted by \cite{gnecco2024extremal}, a lack of samples exceeding the corresponding conditional $\tau$th quantile leads to empirical estimators with large bias and variance. The high dimensionality of the predictor space introduces additional bias, as noise covariates can obscure true signals.

	Current research on extreme quantiles primarily focuses on estimation within fixed-dimensional settings and is either not applicable or less competitive in high-dimensional settings. Although tree-based methods \citep{velthoen2023gradient, gnecco2024extremal} demonstrate effectiveness in relatively large and fixed covariate dimensions, 
	addressing scenarios where the covariate dimension grows with the sample size, comparable to the sample size itself, remains inadequately explored. 
	Recently, \cite{sasaki2024high} proposed a high-dimensional tail index regression model, assuming that the dimension grows with $n$ and is comparable to $n(1-\tau)$. However, this approach imposes constraints such as a Pareto tail and specific link functions between the extreme value index and covariates.
	
	
	In this article, we propose novel estimators for extreme quantiles in high dimensions by integrating concepts from extreme value theory with regularized estimators. Building upon a linear conditional quantile model, we employ regularized quantile regression to estimate intermediate conditional quantiles and extrapolate them to extreme quantiles. Here, ``intermediate quantile" refers to a quantile level $\tau_n$ that approaches one at a moderate rate such that $n(1-\tau_n)\rightarrow\infty$, whereas ``extreme quantile" denotes a quantile level $\tau_n$  that approaches one more rapidly.
	
	{The theoretical analysis of extreme quantiles in high dimensions is challenging. The limiting distribution of a quantile estimator could be intractable in high dimensions, even at central quantile levels. Existing work on central quantiles within $(0,1)$ often requires a common assumption, namely
		$f_{Y}(Q_Y(\tau|\bX)|\bX)>C>0$,
		where $f_{Y}(\cdot|\bX)$ is the conditional density of $Y$ given $\bX$, and $Q_Y(\tau|\bX)$ is the $\tau$th conditional quantile of $Y$; see for instance \citet{belloni2011ℓ1},
		and \citet{wang2012quantile}. However, this assumption is violated in the tails as $\tau$ approaches $1$.
		It is even more challenging to achieve a uniform result of the regularized estimator over the entire tail, which is crucial for constructing an effective extreme value index estimator. 
		This paper addresses the gap and establishes the uniform error rate for quantile regression estimator in high dimensions and at intermediate quantiles. This rate includes terms analogous to those in high-dimensional quantile regression at a central quantile level, augmented by an inflationary component attributed to escalating quantile levels or diminishing effective sample sizes. 
		{In addition, we propose a refined Hill approach for estimating the extreme value index, which is based on a fixed number of intermediate high-dimensional quantile estimates. This framework enables us to analyze the uniform behavior of regularized quantile regression in the tail region and assess the rate at which $f_{Y}(Q_Y(\tau|\bX)|\bX)$ tends to zero. The error rates of the proposed refined Hill and extreme quantile estimators offer theoretical insights into threshold selection in high-dimensional contexts.} 

		The rest of the article is organized as follows. In Section \ref{sec:HEQR}, we present the proposed estimators for the extreme value index and extreme conditional quantiles, and derive theoretical results for the proposed estimators using techniques from both extreme value theory and high-dimensional statistics theory. We assess the finite sample performance of the proposed methods through a simulation study in Section \ref{sec:simul} and the analysis of auto insurance data in Section \ref{sec:application}. Technical details and additional information for Sections \ref{sec:simul} and \ref{sec:application} are provided in the online Supplementary Material.

		
		\section{Extreme Quantile Estimation in High Dimensions}
		\label{sec:HEQR}
		Suppose we observe a random sample $\{(\bX_i, Y_i), i=1, \ldots, n\}$ of the random vector $(\bX, Y)$, where $Y_i$ is the univariate response variable and $\bX_i=(X_{i 1}, \ldots, X_{i p})^T$ is the $p$-dimensional centralized design vector. This article considers the high dimensional case: $p := p(n)\rightarrow \infty$ as $n\rightarrow\infty$. Let $Q_Y(\tau | \bX)=\inf \{y: F_Y(Y |\bX) \geq \tau\}$ denote the $\tau$th conditional quantile of $Y$ given $\bX$, where $F_Y(\cdot | \bX)$ is the cumulative distribution function (CDF) of $Y$ given $\bX$. Denote $\mathcal{X}\subset \mathbb{R}^p$ as the support of $\bX$.
		
		
		Throughout the article, we assume that $F_Y(\cdot | \bX)$ is in the maximum domain of attraction of an extreme value distribution $G_\gamma(\cdot)$ with the extreme value index (EVI) $\gamma>0$, denoted by $F_Y(\cdot | \bX) \in D(G_\gamma)$. That is, for a given random sample $Y_1, \ldots, Y_n$ from $F_Y(\cdot | \bX)$, there exist constants $a_n>0$ and $b_n \in \mathbb{R}$ such that
		$$
		P\left(\left.\frac{\max _{1 \leq i \leq n} Y_i-b_n}{a_n} \leq y\right|\bX\right) \rightarrow G_\gamma(y)=\exp\{-(1+\gamma y)^{-1 / \gamma}\},
		$$
		as $n \rightarrow \infty$, for $1+\gamma y \geq 0$. 
		In this paper, we assume $\gamma>0$, which means that $Y|\bX$ has heavy-tailed distributions as commonly seen in many applications, such as stock market returns, insurance claims, earthquake magnitudes, river flows during floods.

		
		
		The main objective of this paper is to estimate the conditional quantile $Q_Y(\tau_n | \bX)$ when $\tau_n\rightarrow 1$ and $p\rightarrow\infty$ as $n\rightarrow\infty$. We address two distinct regimes for the quantile level $\tau_n$  as it approaches 1: intermediate order quantile levels such that  $n(1-\tau_n)\rightarrow \infty$, and extreme order quantile levels with $n(1-\tau_n)\rightarrow C$, where $C\ge 0$ is some constant. 
		
		
		
		
		We focus on the following tail linear quantile regression model:
		\begin{equation}
			Q_{Y}(\tau| \bX)=\beta_0(\tau)+X_{1}\beta_{1}(\tau)+\cdots+X_{p}\beta_{p}(\tau) =: \bZ^{T}\bbeta(\tau), \quad \text{for all} \ \tau\in [\tau_{ln},1),\label{model1}   
		\end{equation}
		where $\bZ=(1,\bX^T)^T$, $\tau_{ln}\rightarrow 1$ as $n\rightarrow \infty$, and the quantile slope coefficients $\bbeta(\tau)$ may vary across $\tau \in$ $[\tau_{ln}, 1)$. {The linear quantile model in (\ref{model1}) is specifically assumed at the upper tail. Similar tail model specifications have also been considered in \cite{wang2012estimation}, \cite{li2019extreme}, \cite{xu2022prediction}. In high-dimensional settings, we extend this assumption and additionally impose a sparsity condition to ensure model identifiability.}
		We assume there are $s$ tail-relevant variables, that is, $s:=\#\{ j=1,2,\ldots,p: \beta_{j}\left(\tau\right)\neq 0,$
		$\exists \tau\in [\tau_{ln}, 1)\}$, and $s=o(n)$. {A tail-relevant variable may influence the tail quantiles of $Y|\bX$, even if it does not affect the central quantiles.} 
	Without loss of generality, we assume that the first $s$ slope coefficients are nonzero, i.e., $\bbeta(\tau)=(\beta_0(\tau),\beta_1(\tau),\ldots,\beta_s(\tau),\mathbf{0}_{p-s})^{T}$. 

	We propose a three-step procedure to estimate the extreme conditional quantile. In the first step, we obtain $\ell_1$-penalized quantile estimators at a sequence of intermediate quantile levels. For heavy-tailed distributions, the Hill estimator \citep{hill1975simple} is the most commonly used method to estimate the EVI. 
	However, based on the average of log excesses, the Hill estimator requires analysis of the tail quantile process \citep{de2007extreme}, which poses both computational and theoretical challenges in high dimensions. To address these challenges, we develop a refined Hill estimator in the second step, which relies on quantile estimates at a fixed number of intermediate quantile levels that approach to one at the same rate. In the third step, we develop an extrapolation estimator for the extreme conditional quantile $Q_Y(\tau_n| \bX)$ by leveraging extreme value theory and results from the first two steps.

	Before presenting the proposed estimators and their theoretical properties, we first introduce the notations used throughout the paper.  For a sequence of random variables $E_1,E_2,\ldots,E_n$, we denote its order statistics as $E_{(1)}\le E_{(2)}\le\ldots\le E_{(n)}$.   Given a random sample $\bZ_1, \ldots, \bZ_n$, let $\mathbb{G}_n(f)=\mathbb{G}_n\{f(\bZ_i)\}:=n^{-1 / 2} \sum_{i=1}^n[f(\bZ_i)-\mathrm{E}\{f(\bZ_i)\}]$ and $\mathbb{E}_n f=\mathbb{E}_n f\left(\bZ_i\right):=n^{-1}\sum_{i=1}^n f(\bZ_i) $. 
	We denote the $\ell_2$, $\ell_1$,  $\ell_{\infty}$ and $\ell_0$ norms by $\|\cdot\|$, $\|\cdot\|_1$, $\|\cdot\|_{\infty}$ and $\|\cdot\|_0$, respectively.  Let $\|\bbeta\|_{1, n}=\sum_{j=1}^p \widehat{\sigma}_j|\beta_j|$ denote the weighted $\ell_1$-norm with $\widehat{\sigma}_j^2:=\mathbb{E}_n(X_{ij}^2)$.  Given a vector $\bdelta \in \mathbb{R}^p$, and a set of indices $T \subset\{1, \ldots, p\}$, we denote by $\bdelta_T$ the vector where $\delta_{T j}=\delta_j$ if $j \in T, \delta_{T j}=0$ if $j \notin T$.  We use $a \lesssim b$ to denote $a=O(b)$, meaning $a \leq c b$ for some constant $c>0$ that does not depend on $n$; and $a  \asymp b$ to denote $a=O(b)$ and $b=O(a)$. We use $a  \lesssim_{P} b$ to denote $a=O_P(b)$. Additionally, we use $a \vee b=\max \{a, b\}$ and $a \wedge b=\min \{a, b\}$. For notational simplicity, let $F_i(\cdot)=F_Y(\cdot | \bX_i)$ be the conditional distribution function of $Y$ given $\mathbf{X}_i$, and denote $f_i(\cdot)=f_{Y}(\cdot | \bX_i)$.
	Finally, let $\mathbb{X}=(\mathbf{X}_1, \ldots, \mathbf{X}_n )^{\top}, \mathbb{Z}=(\mathbf{Z}_1, \ldots, \mathbf{Z}_n)^{\top}$, and $\mathbf{Y}=(Y_1,Y_2,\ldots, Y_n)^T$. 

	\subsection{$\ell_1$-penalized Intermediate Quantile Estimator}
	\label{subsec:ell1}
	
	Define $\mathcal{T}_n:=\{c(1-\tau_{0n}):c\in [c_1,c_2]\},$ where 
	$\tau_{0n}$ is  an intermediate quantile level such that $\tau_{0n}\rightarrow 1$ and $n(1-\tau_{0n})\rightarrow \infty$, and  $0<c_1< c_2<\infty$ are constants. 
	For any $\tau$ such that $1-\tau\in \mathcal{T}_n$, we define the $\ell_1$-penalized quantile estimator of $\bbeta(\tau)$ as
	\begin{equation}
		\label{ell1QR}
		\hatbbeta(\tau)= \argmin_{(\beta_{0},\bbeta^T)^T\in \mathbb{R}^{p+1}}  \frac{1}{n}\sum_{i=1}^n \rho_{\tau} (Y_i-\beta_{0}-\bX_i^T\bbeta) +\frac{\lambda\sqrt{\tau(1-\tau)}}{n} \|\bbeta\|_{1,n},
	\end{equation}
	{where $\lambda>0$ is the penalization parameter.}

	For high dimensional settings, the choice of $\lambda$ is crucial for achieving estimation consistency. As suggested by \cite{bickel2009simultaneous}, $\lambda$ should exceed a suitably rescaled (sub)gradient of the sample criterion function evaluated at the true parameter value. While the asymptotic scale of $\lambda$ has been derived for central quantiles \citep{belloni2011ℓ1, zheng2013adaptive}, this scale differs for intermediate quantiles as $\tau$ approaches one. {Specifically, a subgradient of the penalized quantile regression objective function at $\bbeta(\tau)$ is given by \(\mathbb{E}_n[\bZ_i \{\tau - \mathbbm{1}(Y_i \leq \bZ_i^T \bbeta(\tau))\} ]+{\lambda\sqrt{\tau(1-\tau)}}n^{-1}(0,\mathbf{g}^T)^T\), where $\bZ=(1,\bX^T)^T$ and $\mathbf{g}=(g_1,\ldots, g_p)^T$ is a subgradient of $\|\bbeta\|_{1,n}$ with $|g_j|=\widehat{\sigma}_j$ for $j=1,2,\ldots,p$. The infinity norm of the first term is of the same order for both central and intermediate quantiles; however, the second term varies in order due to the factor \(\sqrt{\tau(1 - \tau)}\).} Lemma 1 in the 
	Supplementary Material provides an asymptotic uniform lower bound of $\lambda$ over $\tau\in\mathcal{T}_n$, and a practical choice for $\lambda$ is discussed in Section \ref{Subsec:Computation}.

	One major challenge in establishing the theoretical properties of $\hatbbeta(\tau)$ arise from the violation of the condition $f_{Y}(Q_Y(\tau|\bX)|\bX)>C>0$ at intermediate quantiles. 
	This violation complicates the establishment of a quadratic lower bound for the Taylor expansion of the expected quantile loss function. 
	Consequently, the general framework of \cite{negahban2012unified} cannot be applied to obtain the convergence rate through the restricted strong convexity property of the empirical loss function. However, leveraging condition C5 and extreme value theory, we assess the rate at which the conditional quantile and density converge, determining how $f_{Y}(Q_Y(\tau|\bX)|\bX)$ approaches zero as $\tau$ goes to one. This allows us to establish an asymptotic quadratic lower bound for the expected intermediate quantile loss after appropriate standardization (see Lemma 4 in the 
	Supplementary Material).
	{
		To establish the theoretical properties of $\widehat\bbeta(\tau)$ over $\tau\in\mathcal T_n$, we impose Conditions C1-C6. For brevity, the conditions are detailed in Supplementary Material, but we highlight one key condition, C5, below.}

	
	\noindent \bcon C5.
	\label{con6} 
	There exists an auxiliary line $\bZ \rightarrow \bZ^{T} \boldsymbol{\theta}_r$ with $0<r<1$ and a bounded vector $\boldsymbol{\theta}_r$ such that for $U=Y-\bZ^T \boldsymbol{\theta}_r$ and some heavy-tailed distribution function $F_0(\cdot)$ with extreme value index $\gamma>0$ and density $f_0(\cdot)$, the following hold for some positive continuous functions $K(\cdot),{K}_1(\cdot),{K}_2(\cdot)$ on $\mathcal{Z}=(1,\mathcal{X})$.
	\begin{enumerate}[(i)]
		\item There exists positive sequences $d_n,d_{1n}$ such that $ K(\bZ)\asymp d_n,K_1(\bZ)\asymp d_{1n}$ hold for all $\bZ\in\mathcal{Z}$.
		\item For any sequence $t_n\rightarrow\infty$, if $d_n \{1-F_0(t_n)\}\rightarrow0$ and  $d_n  f_0(t_n)\rightarrow0$ as $n\rightarrow\infty$, then 
		$$\begin{aligned}
			\left|\frac{1-F_{U}(t_n | \bZ)}{K(\bZ)\{1-F_0(t_n)\}}-1\right|&=\{1-F_0(t_n)\}^\delta {K}_1(\bZ)\{1+o(1)\},~\text{and}\\
			\left|\frac{f_{U}(t_n | \bZ)}{K(\bZ)f_0(t_n)}-1\right|&=\{f_0(t_n)\}^\delta {K}_2(\bZ)\{1+o(1)\},
		\end{aligned}$$
		holds uniformly for $\bZ\in\mathcal{Z}$, where $\delta>0$ is a constant.
		\item 
		$U_0(t):=F_0^{-1}(1-1 / t)$ satisfies the second-order condition
		$
		A_1(t)^{-1}\{U_0(t z )/U_0(t )-z^\gamma\} \rightarrow z^\gamma(z^{\rho}-1) / \rho, \text { as } t \rightarrow \infty$	
		with $\gamma>0, \rho<0$, and $A_1(t)=\gamma d t^{\rho}$ with $d \neq 0$. Additionally, $f_0(t)$ is regularly varying at infinity with index $-1/\gamma-1$.
	\end{enumerate}
	
	\begin{remark}
		{
			Condition C5 presents a novel framework tailored for high-dimensional settings, where 
			$K(\bZ)$, $K_1(\bZ)$, and $K_2(\bZ)$ may be unbounded and affect the tail behavior of $Y|\bX$. Unlike fixed-dimensional cases \citep{wang2012estimation, chernozhukov2005extremal}, this condition addresses the combined complexities of high dimensionality and tail behavior.
			The conditions 
			$d_n\{1-F_0(t_n)\}\rightarrow0$ and $d_n f_0(t_n)\rightarrow0$ as $n\rightarrow\infty$ ensure that the tail effect of $1-F_0(t_n)$ outweighs the diverging scale effect of $K(\bZ)$ in high dimensions.  
			This requirement allows us to apply extreme value theory to infer the tail behavior of $Y$, and is mild, as 
			it is automatically satisfied when the number of important variables $s$ is finite (see Example 1).  {The second-order condition of \(U_0\) implies that
				${U_0(tz)}/{U_0(t)} \rightarrow z^\gamma$,
				which can be interpreted as a first-order condition (see Condition C6 for the definition). The first-order condition is a necessary and sufficient condition for \(F_0\) to belong to the maximum domain of attraction \(D(G_\gamma)\) for heavy-tailed distributions \citep{de2007extreme}. Most commonly used families of continuous distributions satisfies the second-order condition. Although both the extreme value index estimator and the extreme quantile estimator are based on the first-order condition, the second-order condition is essential for deriving their theoretical results.
			}Additionally, Lemma 3 in the Supplementary Material demonstrates that under Condition C5, a second-order condition of $F_{Y}(t|\bX)$ still holds, similar to the fixed-dimensional case.
		}  
	\end{remark}

	{Define $s_r:=\|\btheta_r\|_0$. The following Theorem \ref{theorem:interQR}  establishes a uniform bound of $\widehat{\bbeta}(\tau)-\bbeta(\tau)$ for $\tau\in\mathcal{T}_n$.}

	\begin{theorem}
		\label{theorem:interQR}
		Assume Conditions C1-C6 hold, along with $1-\tau_{0n}>\sqrt{s\log (p)/n}$, $d_n^{-1}d_{1n}^{1/\delta}(1-\tau_{0n})\rightarrow 0$, $s_r d_n^{-\gamma}(1-\tau_{0n})^{\gamma}\rightarrow 0$, and $\lambda\asymp \sqrt{n \log p}/\sqrt{1-\tau_{0n}}$. Then, for any $\epsilon\in (0,1)$ and some positive constant $C$, there exist a positive integer $N$ such that, for all $n>N$,
		$$
		\sup _{1-\tau \in \mathcal{T}_n }\|\widehat{\bbeta}(\tau)-\bbeta(\tau)\| \leq C d_n^{\gamma}(1-\tau_{0n})^{-\gamma-1}\sqrt{\frac{s\log (p\vee n)}{n}},
		$$ with probability at least $1-4 \epsilon-5 p^{-4}$.
	\end{theorem}

	Theorem \ref{theorem:interQR} establishes bounds  similar in spirit to those found in \cite{chernozhukov2005extremal} but with an additional term,  $d_n^{\gamma}\left(1-\tau_{0n}\right)^{-\gamma-1}$, which inflates the upper bound. This term reflects the decay of the conditional density in the tails. Specifically, as shown in Remark S.2 in the Supplementary Material, under certain conditions, as $n\rightarrow\infty$,
	$f_{Y}(Q_Y(\tau|\bX)|\bX)\sim d_n^{-\gamma}(1-\tau_{0n})^{\gamma+1},$
	which describes the rate at which the conditional density approaches zero at intermediate quantiles.
	Theorem 1 poses both upper and lower bounds conditions on \(1 - \tau_{0n}\). The condition $1-\tau_{0n}>\sqrt{s\log (p)/n}$ prevents $\tau_{0n}$ from approaching one too quickly, aligning with the requirement for restricted strong convexity of the objective function (see Lemma 4 in the Supplementary Material). Restricted strong convexity refers to maintaining strong convexity within a restricted set \citep{negahban2012unified}. 
	In the high-dimensional context, this restricted set is defined such that the error vector lies within it with high probability, given a suitably chosen penalty and certain conditions. Intuitively, restricted strong convexity holds only if the prediction error of the penalized estimator, $\bZ^T(\hatbbeta(\tau)-\bbeta(\tau))$, grows slower than the quantile of $Y|\bX$. The conditions $s_r d_n^{-\gamma}(1-\tau_{0n})^{\gamma}\rightarrow 0$ and $d_n^{-1}d_{1n}^{1/\delta}(1-\tau_{0n})\rightarrow 0$ ensure that the tail effect dominates the high-dimensional effect. In Section \ref{subsec:EQ}, we provide an example illustrating the feasibility of these conditions.

	\subsection{Refined Hill Estimator}
	\label{subsec:RHE}
	Let $k=\lfloor n(1-\tau_{0n})\rfloor$. For simplicity, we omit the rounding notation, as it does not impact the theoretical discussion that follows. Define quantile levels $\tau_{0n}=\tau_{1}<\tau_{2}<\cdots<\tau_{J} \in(0,1)$, where $\tau_j=1-l_jk/n, l_j=s^{j-1}, s\in(0,1)$. For each $j=1,2,\ldots, J$, we estimate $\bbeta(\tau_j)$ by the $\ell_1$-penalized quantile regression estimator in \eqref{ell1QR}.
	Suppose that we are interested in estimating the extreme conditional quantile of $Y$ given $\bX=\bx \in \mathcal{X}$. 
	Let $\bz=(1,\bx^T)^T$, and we can define the refined Hill estimator as
	$$\hat{\gamma}=\left(\sum_{j=1}^J \phi\left(l_j\right) \log \left(1 / l_j\right)\right)^{-1} \sum_{j=1}^J \phi\left(l_j\right)\log \left( \frac{\bz^T \hatbbeta(\tau_{j})}{\bz^T \hatbbeta(\tau_{1})}\right).
	$$
	where $\phi(\cdot) $ is some positive measurable function. 
	
	
	
	
	

	{
		The refined Hill estimator is a weighted variant of the Hill-type estimators found in the literature but with the following key differences. First, \(\hat{\gamma}\) utilizes a fixed number of log excesses at intermediate quantile levels with the same rate, whereas the Hill estimator typically relies on quantiles approaching one at varying rates, as seen in \citep{wang2012estimation}. Second, \(\hat{\gamma}\) allows for flexible weighting through \(\phi\), while the Hill estimator commonly applies equal weights to upper quantiles \citep{hill1975simple, wang2012estimation, daouia2023inference}. We will discuss the choice of the weight function $\phi$, and the tuning parameters $J$ and $s$ in Section \ref{Subsec:Computation}. Similar weighting schemes have been explored by other researchers to generalize Hill and Pickands estimators \citep{drees1995refined, daouia2013kernel,  he2022extremal} and for estimating the Weibull tail coefficient \citep{gardes2016estimation, he2020extremalWeibull}. These methods were developed in fixed-dimensional settings where the theoretical guarantees of the refined estimators could be established using the tractable asymptotic normality of intermediate quantile estimates. While this strategy is not feasible in high-dimensional settings, we can leverage the uniform convergence results from Theorem 1 to establish the convergence rate of the refined Hill estimator. Suppose there exist a positive sequence $d_{1n}$ such that, $ K_1(\bZ)\asymp d_{1n}$ hold for all $\bZ\in\mathcal{Z}$.}

	\begin{theorem}
		\label{Theorem:RHgamma}
		Assume that $\bx$ belongs to a compact set in $\mathcal{X}\subset\mathbb{R}^p$.
		Assume the conditions of Theorem \ref{theorem:interQR}, and that $s_r=o\left(\left(d_n^{\gamma+\rho}(n/k)^{\gamma+\rho}\right) \vee \left(d_n^{\gamma-\delta} d_{1n} (n/k)^{\gamma-\delta}\right)\right)$ , $ d_n^{\rho}(n/k)^{-\rho}\vee d_{1n}d_n^{-\delta}(n/k)^{\delta}=o\left((ns/k)\sqrt{{\log (p\vee n)}/n}\right)$.  Then we have
		$$\hat{\gamma}-\gamma=O_p\left(\frac{ns}{k}\sqrt{\frac{\log (p\vee n)}{n}}\right).$$
	\end{theorem}
	

	
	\begin{remark}
		Theorem \ref{Theorem:RHgamma} suggests that if $(ns/k)\sqrt{{\log (p\vee n)}/n}\rightarrow 0$, then $\hat{\gamma} \overset{p}{\rightarrow } \gamma$. In Theorem \ref{Theorem:RHgamma}, the conditions on $s_r$ are posed to account for the error in the second-order term in the tail expansion of $F_Y(\cdot|\bX)$. These conditions can be simplified for specific models, as discussed in Section \ref{subsec:EQ}. 
		Note that the convergence rate of $\hat\gamma$ is $(ns/k)\sqrt{\log (p\vee n)/n}$, which is slower than the typical ${k}^{-1/2}$ rate achieved in the fixed-dimensional case \citep{wang2009tail,daouia2013kernel}. The difference arises from the slower convergence of the intermediate quantile estimators in high-dimensional settings. Furthermore, to ensure consistency, $k$ must satisfy $k > n^{1/2}$,  which is stricter than the requirement in the fixed dimensional case \citep{wang2012estimation}, where $k>n^{\eta}$ with $\eta$ being a small positive constant. Finally, while the error bound of the \(\ell_1\)-penalized intermediate quantile estimator depends on \(\gamma\), the convergence rate of $\hat\gamma$ remains invariant across \(\gamma\).
	\end{remark}
	

	\subsection{Extreme Conditional Quantile Estimator}
	\label{subsec:EQ}
	Let $\tau_n\rightarrow 1$ be an extreme quantile level such that
	$1-\tau_n=o(1-\tau_{0n})$. 
	Suppose that we are interested in estimating the $\tau_n$th conditional quantile of $Y$ given $\bX=\bx$. 
	Direct estimation using the $\ell_1$-penalized quantile estimator in  (\ref{ell1QR}) is often inaccurate or unstable due to limited data in the extreme tail. However, we can leverage extreme value theory and extrapolate from intermediate to extreme quantiles by using the relationship between quantiles with levels approaching one at different rates.

	
	Define $U_Y(t|\bX)=\inf\{y:F_Y(y|\bX)\ge 1-1/t\}=F^{-1}_Y(1-1/t|\bX)$ for $t\ge 1$, the ($1-1/t$)th quantile of $F_Y(\cdot|\bX)$.
	According to Corollary 1.2.10 in \cite{de2007extreme}, for a heavy-tailed distribution $F_Y(\cdot|\bX=\bx)\in D(G_{\gamma})$, we have
	$$\frac{U_Y(tz|\bx)}{U_Y(t|\bx)}\rightarrow z^{\gamma},\quad \text{as }t\rightarrow\infty.$$
	Motivated by this, for a given $\bx$, we estimate $Q_Y(\tau_n | \bx)$ by
	$$
	\widehat{Q}_Y(\tau_n |\bx)=\left(\frac{1-\tau_{0n}}{1-\tau_n}\right)^{\hat{\gamma}} \bz^T \hatbbeta(\tau_{0n}),
	$$
	where $\bz=(1,\bx^T)^T$.


	\begin{theorem}
		\label{Theorem:EQ}
		Assume the conditions of {Theorem \ref{Theorem:RHgamma}}. Let $\tau_n\rightarrow 1$ be a quantile level such that
		$n(1-\tau_n)=o(k)$, then we have
		$$\frac{\widehat{Q}_{Y}\left(\tau_n|\bx\right)}{{Q}_{Y}\left(\tau_n|\bx\right)}-1=O_p\left(\log\left(\frac{k}{n(1-\tau_n)}\right)\frac{ns}{k}\sqrt{\frac{\log (p\vee n)}{n}}\right).$$
		Additionally, if $\log[k/\{n(1-\tau_n)\}]$ $ (ns/k)\sqrt{{\log (p\vee n)}/{n}}\rightarrow 0$, ${\widehat{Q}_{Y}\left(\tau_n|\bx\right)}/{{Q}_{Y}\left(\tau_n|\bx\right)}\overset{p}{\rightarrow }1$.
	\end{theorem}


	According to our estimation procedure, the error in estimating extreme quantiles primarily arises from the EVI estimator and the \(\ell_1\)-penalized intermediate quantile estimator. Specifically, the error rate of the EVI estimator and the relative error rate of the intermediate quantile estimator are both given by \((ns/k)\sqrt{{ \log(p \vee n)}/{n}}\). Since the EVI estimator contributes to the error rate at the exponent, the final error rate for the extreme quantile estimation is $\log[{k}/\{n(1-\tau_n)\}(ns/k)\sqrt{{\log (p\vee n)}/{n}}$. In the fixed dimension case, the rate of the relative error of extreme quantile estimator is $\log[k/\{n(1-\tau_n)\}]/\sqrt{k}$. Comparing this with our results, we observe that the inflated term arises from the EVI estimator, which has an error rate of \({1}/{\sqrt{k}}\) in the fixed-dimensional setting.

	\begin{remark}
		In this paper, we assume that \( F_Y(\cdot | \bX) \in D(G_{\gamma})\), which represents a first-order condition on \( F_Y(\cdot | \bX) \). To derive the properties of \( \widehat{Q}_Y(\tau_n | \bx) \), a second-order condition is needed. Condition C5 assumes the existence of an auxiliary line so that $U=Y-\bZ^T\btheta_r$ satisfies the second-order condition. Under this condition, we can show that $F_Y(\cdot | \bX)$ also satisfies the second-order condition even for high-dimensional $\bX$ (see Lemma 3), and it has the same EVI across  \(\bX \in \mathcal{X}\). This implies that the EVI can be technically estimated based on the intermediate conditional quantiles at any \(\bx \in \mathcal{X}\). 
		Theorem \ref{Theorem:RHgamma} establishes the convergence rate for $\hat{\gamma}$ obtained at $\bX = \bx$ belonging to a compact set, which we now denote as $\hat{\gamma}(\bx)$ to emphasize its dependence on $\bx$.
		Careful selection of \(\bx\) is important, particularly in high-dimensional cases where data points may be sparse due to the curse of dimensionality. In practice, we recommend estimating \(\hat{\gamma}(\bx)\) at \(\bar \bx = n^{-1} \sum_{i=1}^n \bX_i\), as this leads to more stable numerical performance and a higher convergence rate than $\hat\gamma(\bx)$, whose convergence rate is presented in Remark 2. Specifically, $\hat{\gamma}(\bar \bx)-\gamma=O_p\left(({n}/k)\sqrt{{s\log (p\vee n)}/{n}}\right)$; see the proof of Theorem \ref{Theorem:RHgamma} in the Supplementary Material for more details.
		A pooled estimator that averages  $\{\hat\gamma(\bX_i), i=1,\ldots,n\}$ could also improve estimation efficiency; however, this approach is computationally intensive and less practical in high-dimensional settings.
		
	\end{remark}
	

	The regularity conditions outlined in our paper encompass a broad range of conventional regression settings. To illustrate these conditions, we consider the location-scale shift model as a representative example.
	
	\noindent {\bf Example 1.} Consider the location-scale shift model:
	$$Y=\alpha+\bX^T \boldsymbol{\beta}+(1+\bX^T \widetilde{\boldsymbol{\sigma}}) \varepsilon,$$
	where $1+\bX^T \widetilde{\boldsymbol{\sigma}}>0$ for $\bX \in \mathcal{X}$, and $\varepsilon \sim F_0(\cdot)$ with $F_0$ satisfying the second order condition for some $\gamma>0$ and $\rho<0$. Obviously, $Q_Y(\tau|\bX)=(\alpha+F_0^{-1}(\tau))+\bX^T(\bbeta+\tilde{\boldsymbol{\sigma}}F_0^{-1}(\tau))$. 
	
	Suppose that the scale effect variable is finite, i.e., $\|\tilde{\sigma}\|_0<\infty$. Under this model, Condition C5 holds with $K\left(\bZ\right)=\left(1+\bX^T \widetilde{\boldsymbol{\sigma}}\right)^{{1}/{\gamma}}$, ${K}_1\left(\bZ\right)=C\left\{\left(1+\bX^T \widetilde{\boldsymbol{\sigma}}\right)^{-{\rho}/{\gamma}}-1\right\}$, where $C$ is some constant and $\delta=-\rho$. Details of the derivation can be found in the Supplementary Material. Let $\|\bbeta\|_0=n^{b}$, where $0\le b<1$. Then $d_n\sim  C$, $d_{1n}\sim C$, and $s_r=n^{b}$. Thus the conditions of {Theorem \ref{Theorem:RHgamma}} are  $k\rightarrow\infty$, $k/n\rightarrow 0$, $\sqrt{s \log p} \sqrt{n} \lesssim k$, $n^b (k/n)^{\gamma}\rightarrow 0$,  for any $\gamma>0$, which implies  
	$\sqrt{s\log p}\sqrt{n}\lesssim k<\min\{n s_r^{-1/\gamma},n\}.$
	The error rate  is $\sup _{1-\tau \in \mathcal{T}_n }\|\widehat{\bbeta}(\tau)-\bbeta(\tau)\|=O_p\left((n/k)^{\gamma+1}\sqrt{s\log (p\vee n)/n}\right)$. Next, we consider two special cases to simplify the conditions on $k$ and discuss the feasibility.
	\medskip
	\begin{itemize}
		\item [{\bf (I).}] The non-zero components of \(\bbeta\) and \(\bbeta(\tau)\) are at the same positions, i.e., \(s = s_r = n^b\), where {$0<b<1$}.
		
		{Theorem \ref{theorem:interQR} requires the condition \(\sqrt{s \log p} \sqrt{n} \lesssim k < n s^{-{1}/{\gamma}}\).
			The addtional conditions on $k$ in Theorems \ref{Theorem:RHgamma} and \ref{Theorem:EQ} are \(d_n^{\rho}(k/n)^{-\rho} \vee d_{1n}d_n^{-\delta}(k/n)^{\delta} = o((ns/k) \sqrt{\log (p \vee n) / n})\) and \(s_r = o\left((d_n^{\gamma+\rho}(n/k)^{\gamma+\rho}) \vee (d_n^{\gamma-\delta}d_{1n}(n/k)^{\gamma-\delta})\right)\), which can be simplified to \(k \lesssim (s^2 \log p)^{1/2(1-\rho)} n^{1 - 1/2(1-\rho)}\) and \(k \lesssim n s^{-{1}/{(\gamma+\rho)}}\), respectively. To obtain consistency of $\hat{\gamma}$, the condition on $k$ is $s\sqrt{\log p} \sqrt{n}\lesssim k$.
			In summary, the condition on \(k\) is 
			$
			s\sqrt{ \log p} \sqrt{n} \lesssim k \lesssim \left\{(s^2 \log p)^{1/2(1-\rho)} n^{1 - 1/2(1-\rho)}\right\} \wedge n s^{-1/(\gamma+\rho)}.
			$
			To ensure the existence of an appropriate \(k\), the additional sparsity conditions are 
			$s^{1-1/(2\rho)}\log(p)/n\rightarrow 0$ and $s^{1+{2}/{(\gamma+\rho)}}\log p/n\rightarrow 0$.}

		

		
		
		\item[{\bf (II).}] Suppose that \(\|\bbeta\|_0\) is finite, that is, \(b=0\). 
		
		In this case, the condition on \(k\) becomes less stringent. Specifically, the condition in Theorem 1 simplifies to \(\sqrt{\log p} \sqrt{n} \lesssim k < n\). Furthermore, the conditions on \(k\) in Theorems \ref{Theorem:RHgamma} and \ref{Theorem:EQ} can be expressed as \(\sqrt{\log p} \sqrt{n} \lesssim k \lesssim (\log p)^{1/2(1-\rho)} n^{1 - 1/2(1-\rho)}\). These conditions are feasible for any \(\rho < 0\).
		
		
	\end{itemize}
	
	
	\subsection{Computational Issues}
	\label{Subsec:Computation}

	The proposed method relies on several tuning parameters/functions, including the penalization parameter $\lambda$, the integer $k$ associated with the intermediate quantile level $\tau_{0n}=1-k/n$, the weight function $\phi(\cdot)$ and the number of intermediate quantile levels $J$ in the refined Hill estimator. In this section, we will discuss practical methods for selecting these quantities. 
	
	\smallskip
	
	{
		\noindent{\bf Selection of $\lambda$.}  In high-dimension settings,  the optimal choice of \(\lambda\) depends on many factors, including $p$, the design matrix and error distribution. \cite{bickel2009simultaneous} suggested that \(\lambda\) should exceed the supremum norm of a suitably rescaled (sub)gradient of the sample criterion function, \cite{belloni2011ℓ1} proposed a similar approach for selecting $\lambda$ at central quantiles. However, applying this method directly to intermediate quantiles often results in over-shrinkage.}
	{Our theoretical study suggests that at intermediate quantiles, $\lambda$ should scale as \(C \sqrt{n \log p} / \sqrt{1-\tau_{0n}}\), but accurately estimating the constant \(C\) is challenging, as noted in \cite{homrighausen2017risk, chetverikov2021cross}. In our implementation, we conduct 10-fold cross-validations to select $\lambda$ for each $\tau_j, j=1,2,\ldots, J$. This involves splitting the data into ten folds and using each fold to evaluate the estimator based on the remaining folds. For each \(\lambda\), we calculate the average quantile loss at $\tau_{j}$ across all folds and select the \(\lambda\) that minimizes this loss.}
	{While time-intensive, our numerical studies show that cross-validation performs well across various settings considered. Developing an efficient and data-driven  selection method tailored to intermediate quantiles is an interesting direction for future research.} 
	

	\smallskip
	
	\noindent{\bf Selection of $k$.} {The intermediate quantile level, or equivalently \( k \), plays an important role of balancing between bias and variance. A small \( k \) increases variance, while a large $k$ leads to higher bias. A common approach is to select \( k \) at the first stable point in the plot of the EVI estimator versus \( k \); see \cite{neves2015modeling} for a heuristic algorithm. However, this method is computationally intensive in high-dimensional cases, as the estimator needs to be calculated over a sequence of \( k \) values,  and the EVI may exhibit local variations across $\bx$. To reduce the computational cost, we propose a rule of thumb for selecting $k$. Based on Theorem 2 and Example 1, \( k \) should satisfy  \(\sqrt{s\log p}\sqrt{n} \lesssim k < \min\{n s_r^{-1/\gamma}, n\}\). Motivated by this, we set \( k = \lfloor c_0 n^{0.5+\delta_1} (\log p)^{0.5+\delta_2} \rfloor \), where $c_0>0$ is a constant, and $\delta_1$ and $\delta_2$ are small positive constants. The sensitivity analysis in the Supplementary Material shows that our proposed method remains stable for \( c_0 \in [0.5, 2.3] \), \( \delta_1 \in [0.005, 0.016] \), and \( \delta_2 \in [0.01, 0.08] \), across various scenarios and sample sizes. Throughout our numerical studies, we use $c_0=0.8$, $\delta_1=0.01$ and $\delta_2=0.05$.}



\smallskip

{
	\noindent {\bf Choice of weights.} The refined Hill estimator assigns different weights to $J$ intermediate quantiles. In our implementation, we use \(\phi(\ell_j) = \ell_j^a\) and \(\ell_j = s^{j-1}\), where \(a, s \in (0,1)\), as recommended by \cite{he2022extremal} for extreme analysis in an autoregressive model. The authors showed that, in their setup, the asymptotic variance of the refined Hill estimator is proportional to a term depending only on \((J, s, a)\).  Minimizing this term yields the optimal choice \((J, s, a)\). While the limiting distribution is intractable without debiasing in high-dimensional settings, we find that the same parameters continue to perform well, resulting in more efficient estimation than methods based on the unweighted Hill estimator.
}



\section{Simulation Study}
\label{sec:simul}

{
	We conduct a simulation study  to investigate the performance of the proposed method for estimating extreme quantiles in high-dimensional settings. The data are generated from
	$$Y_i=X_{i1}+X_{i2}+ (1+wX_{i1})\varepsilon_i,\qquad i=1,2,\ldots,n,$$
	where $X_{ij}\sim U(0,1)$ for $j=1,2,\ldots,p$ and $\varepsilon_i$ are independent and identically distributed student t random variables, and $w$ is a constant controlling the degree of heteroscedasticity. Under this model, the $\tau$th conditional quantile of $Y$ is 
	$Q_Y(\tau|\bX_i)=\alpha(\tau)+\bX_i^T\bbeta(\tau),$
	where $\bX_i=\left(X_{i1},X_{i2},\ldots,X_{ip}\right)^T$, $\alpha(\tau)=Q_{\varepsilon}(\tau)$, $\bbeta(\tau)=\left(1+wQ_{\varepsilon}(\tau),1,0\ldots,0\right)^T$, and $Q_{\varepsilon}(\tau)$ is the $\tau$th quantile of $\varepsilon_i$. We consider six different cases. In Cases 1, 3, and 5, $\varepsilon_i\sim t(5)$ and in Cases 2, 4 and 6, $\varepsilon_i\sim t(2)$.
	Cases 1-2 have homoscedastic errors with $w=0$, while $w=0.5$ in Cases 3-4, and $w=0.9$ in Cases 5-6. Note that, for $t(v)$, the EVI is $\gamma=1/v$.
	We consider two settings for the sample size $n$ and dimension $p$:
	(1) $p=p(n)$ grows with $n$ but $p$ is smaller than $n$, with $p= \lceil n^{0.5}\rceil$ and $n=1000,5000$; (2) $p$ is larger than or equal to $n$, with $(n,p)=(1000,1000)$ and $(1000,1200)$.
	For each scenario, the simulation is repeated $500$ times.}

{
	For each simulation, we estimate $Q_Y(\tau_n|\bx)$ at $\tau_n=0.995, 0.999$ using three methods: the extreme quantile regression method (EQR) without assuming common slope in \cite{wang2012estimation}; the high-dimensional quantile regression (HQR) of \cite{belloni2011ℓ1}, and the proposed high-dimensional extreme quantile regression (HEQR). For EQR, the number of upper-order statistics $k$ is set to $[4.5n^{1/3}]$, as recommended in \cite{wang2012estimation}.} 

{
For each simulation, we calculate integrated squared error (ISE), defined as
$$
\mathrm{ISE}=\frac{1}{L} \sum_{l=1}^L\left\{\frac{\widehat{Q}_Y(\tau_n | \bX_l^*)}{Q_Y(\tau_n | \bX_l^*)}-1\right\}^2,
$$
where $\bX_1^*, \ldots, \bX_L^*$ are evaluation points of the covariates, and we define the mean integrated squared error (MISE) as the average ISE across $500$ simulations. In our simulation, we set $L=100$, and let $\bX_l^*$ be random replicates of $\bX$.
}

\begin{table}[!htb]
\begin{threeparttable}
	\caption{The mean integrated squared error (MISE) and standard error (in parentheses) of different estimators for the extreme conditional quantile at $\tau_n = 0.995, 0.999$ for $\gamma=0.2$ and $p= \lceil n^{0.5}\rceil$. All values are in percentages.}
	\renewcommand\arraystretch{1.5}
	\label{table:gamma0.2}
	\begin{tabular}{llclll}
		\hline\hline
		\multicolumn{1}{c}{$w$} & \multicolumn{1}{c}{$n$} & $\tau_n$ &\multicolumn{1}{c}{HEQR} & \multicolumn{1}{c}{EQR} & \multicolumn{1}{c}{HQR} \\\hline
		0                       & 1000                    & 0.995                  & 2.9(0.05)                 & 3.6(0.05)               & 5.7(0.13)               \\
		&                         & 0.999                    & 4.7(0.09)                 & 5.2(0.09)               & 12.5(0.12)              \\
		& 5000                    & 0.995                  & 1.6(0.02)                 & 2.7(0.02)               & 3.7(0.04)               \\
		&                         & 0.999                  & 2.4(0.03)                 & 4.2(0.05)               & 6.1(0.06)               \\\hline
		0.5                     & 1000                    & 0.995                   & 3.3(0.05)                 & 4.4(0.06)               & 7.5(0.21)               \\
		&                         & 0.999                   & 4.8(0.10)                 & 5.8(0.10)               & 14.6(0.14)              \\
		& 5000                    & 0.995                    & 2.0(0.02)                 & 3.4(0.03)               & 5.0(0.06)               \\
		&                         & 0.999                    & 2.4(0.03)                 & 4.8(0.06)               & 6.9(0.08)               \\\hline
		0.9                     & 1000                    & 0.995                     & 3.7(0.05)                 & 5.2(0.07)               & 9.0(0.29)               \\
		&                         & 0.999                   & 5.0(0.09)                 & 6.5(0.12)               & 15.8(0.17)              \\
		& 5000                    & 0.995                    & 2.3(0.02)                 & 4.0(0.04)               & 6.0(0.08)               \\
		&                         & 0.999                   & 2.6(0.03)                 & 5.1(0.06)               & 7.8(0.12)              \\\hline\hline
	\end{tabular}
	\begin{tablenotes}
		\tiny
		\item {HEQR}: the proposed estimator; {EQR}: the  method in \cite{wang2012estimation}; {HQR}: the high-dimensional qunatile regression estimator in \cite{belloni2011ℓ1}.
	\end{tablenotes}
\end{threeparttable}
\end{table}
\begin{table}[!htb]
\begin{threeparttable}
	\caption{The mean integrated squared error (MISE) and standard error (in parentheses) of different estimators for the extreme conditional quantile at $\tau_n = 0.995, 0.999$ for $\gamma=0.5$ and $p= \lceil n^{0.5}\rceil$. All values are in percentages.}
	\renewcommand\arraystretch{1.5}
	\label{table:gamma0.5}
	\begin{tabular}{llclll}
		\hline\hline
		\multicolumn{1}{c}{$w$} & \multicolumn{1}{c}{$n$} & $\tau_n$ & \multicolumn{1}{c}{HEQR} & \multicolumn{1}{c}{EQR} & \multicolumn{1}{c}{HQR} \\\hline
		0                       & 1000                    & 0.995                     & 7.4(0.26)                 & 10.7(0.56)              & 24.6(3.33)              \\
		&                         & 0.999                  & 15.0(0.68)                & 19.9(2.11)              & 69.7(16.05)             \\
		& 5000                    & 0.995                  & 5.3(0.11)                 & 7.7(0.12)               & 12.7(0.28)              \\
		&                         & 0.999                   & 7.2(0.23)                 & 10.5(0.26)              & 28.5(2.78)              \\\hline
		0.5                     & 1000                    & 0.995                    & 8.5(0.31)                 & 13.5(0.70)              & 32.3(3.70)              \\
		&                         & 0.999                  & 15.4(0.84)                & 23.8(2.63)              & 96.8(28.76)             \\
		& 5000                    & 0.995                    & 6.3(0.14)                 & 9.7(0.18)               & 17.1(0.42)              \\
		&                         & 0.999                  & 8.1(0.33)                 & 11.9(0.39)              & 36.5(3.93)              \\\hline
		0.9                     & 1000                    & 0.995                    & 9.6(0.33)                 & 16.1(0.80)              & 40.5(4.25)              \\
		&                         & 0.999              & 16.0(0.86)                & 27.3(2.86)              & 119.9(38.54)            \\
		& 5000                    & 0.995                    & 7.3(0.17)                 & 11.5(0.23)              & 21.5(0.58)              \\
		&                         & 0.999                  & 9.4(0.39)                 & 13.5(0.49)              & 46.9(5.11)             
		\\\hline\hline
	\end{tabular}
	\begin{tablenotes}
		\tiny
		\item {HEQR}: the proposed estimator; {EQR}: the  method in \cite{wang2012estimation}; {HQR}: the high-dimensional qunatile regression estimator in \cite{belloni2011ℓ1}.
	\end{tablenotes}
\end{threeparttable}
\end{table}

\begin{table}[!htb]
\begin{threeparttable}
	\caption{The mean integrated squared error (MISE) and standard error (in parentheses) of different estimators for the extreme conditional quantile at $\tau_n = 0.995, 0.999$ for the setting $p\geq n$,
		with $(n, p) = (1000, 1000)$ and $(1000, 1200)$. All values are in percentages.}
\renewcommand\arraystretch{1.5}
\label{table:hd}
\begin{tabular}{lllcccc}
	\hline\hline\multicolumn{1}{c}{$w$} & \multicolumn{1}{c}{$\gamma$} & \multicolumn{1}{c}{$p$} & $\tau_n$     & HEQR     & HQR        \\\hline
	0                       & 0.2                                          & 1000                    & 0.995  & 0.23(0.03) & 0.19(0.02) \\
	&                              &                                             & 0.999   & 0.23(0.03) & 0.43(0.03) \\
	&                                              & 1200                    & 0.995   & 0.20(0.02) & 0.16(0.02) \\
	&                                                 &                         & 0.999   & 0.22(0.03) & 0.30(0.03) \\\hline
	0.5                     & 0.2                                      & 1000                    & 0.995  & 0.26(0.03) & 0.23(0.02) \\
	&                              &                                     & 0.999  & 0.26(0.03) & 0.50(0.03) \\
	&                                     & 1200                    & 0.995  & 0.23(0.03) & 0.20(0.02) \\
	&                                              &                         & 0.999  & 0.24(0.03) & 0.35(0.03) \\\hline
	0                       & 0.5                                  & 1000                    & 0.995  & 0.32(0.04) & 0.44(0.33) \\
	&                                         &                         & 0.999  & 0.41(0.06) & 0.63(0.13) \\
	&                                     & 1200                    & 0.995  &  0.41(0.05) & 0.36(0.23) \\
	&                            &                         & 0.999  & 0.51(0.05) & 0.55(0.14) \\\hline
	0.5                     & 0.5                            & 1000                    & 0.995  & 0.34(0.06) & 0.52(0.71) \\
	&                             &                         & 0.999  & 0.39(0.07) & 0.67(0.17) \\
	&                 & 1200                    & 0.995   & 0.34(0.05) & 0.41(0.41) \\
	&                          &                         & 0.999  & 0.44(0.08) & 0.58(0.20)
	\\
	\hline\hline
\end{tabular}
\begin{tablenotes}
	\tiny
	\item {HEQR}: the proposed estimator; {HQR}: the high-dimensional qunatile regression estimator in \cite{belloni2011ℓ1}.
\end{tablenotes}
\end{threeparttable}
\end{table}

{
We begin by examining the $p= \lceil n^{0.5}\rceil$ setting. Tables \ref{table:gamma0.2} and  \ref{table:gamma0.5} summarize the MISE of three estimators for the extreme conditional quantile at $\tau_n=0.995,$ and $0.999$. 
Results show that for all three methods, both MISE and its standard error increase as $\tau_n$ approaches 1, and as the distribution becomes more heavy-tailed or heteroscedastic.
In all scenarios, the proposed method HEQR outperforms the others in terms of both MISE and standard error.
Both EQR and HEQR rely on extrapolation using extreme value theory. 
In this setting, since $p$ increases with $n$ but remains smaller than $n$, ERQ method is still applicable,
and both EQR and HEQR perform better than HQR. 
The advantages of HEQR over EQR come from its penalized estimation of $\bbeta(\tau)$ and its use of the refined Hill estimator, compared to EQR's unweighted Hill estimator.
}


{
We next examine the setting where $p\geq n$. Table \ref{table:hd} summarizes the results for two methods HEQR and HQR, with $(n, p) = (1000, 1000)$ and $(1000, 1200)$, as EQR is not applicable in these cases. While HQR performs reasonably well for small $\gamma=0.2$ at $\tau_n=0.995$, it is significantly less efficient than HEQR for all other scenarios, regarding both MISE and its standard error. Overall, the results suggest that HEQR demonstrates superior accuracy and stability, particularly for estimating extremely high quantiles in heavy-tailed distributions.}



We perform a sensitivity analysis to assess the stability of the proposed method with respect to the choices of \( c_0 \), \( \delta_1 \), and \( \delta_2 \) in the rule of thumb \( k = \lfloor c_0 n^{0.5+\delta_1} (\log p)^{0.5+\delta_2} \rfloor \). For brevity, we present results only for Case 1. The other cases exhibit similar behavior, and the results are provided in the Supplementary file. Figure \ref{Fig:sensitivityc0} shows the MISE of EQR, HQR and HEQR for the extreme conditional quantiles at \(\tau_n=0.995\) against \(c_0 \in [0.4, 3]\), with \((\delta_1, \delta_2)\) fixed at \((0.01, 0.05)\) for \(n = 1000\) and \(5000\). The shaded area denotes the 95\% pointwise confidence band of the MISE. The two horizontal lines represent the MISE of HQR and EQR, respectively, while the shaded area corresponds to the 95\% pointwise confidence band of the MISE. For \(c_0 \in [0.5, 2.3]\), the MISE of HEQR is consistently smaller than that of HQR and EQR, indicating that \(c_0\) in this range is a suitable choice. Figures \ref{Fig:sensitivitydelta1} and \ref{Fig:sensitivitydelta2} plot the MISE against \(\delta_1 \in [0.005, 0.03]\) and \(\delta_2 \in [0.02, 0.08]\), respectively, with the other two constants fixed at their recommended values. The horizontal line represents the MISE associated with the recommended values \(c_0 = 0.8\), \(\delta_1 = 0.01\), and \(\delta_2 = 0.05\), while the shaded area corresponds to the 95\% pointwise confidence band of the MISE. The results indicate that the method is insensitive to variations in \(\delta_1 \in [0.005, 0.016]\) and \(\delta_2 \in [0.01, 0.08]\).

\begin{figure}
\centering  
\subfigure[$n=1000$]{
	\label{Fig.c01}
	\includegraphics[width=0.45\textwidth]{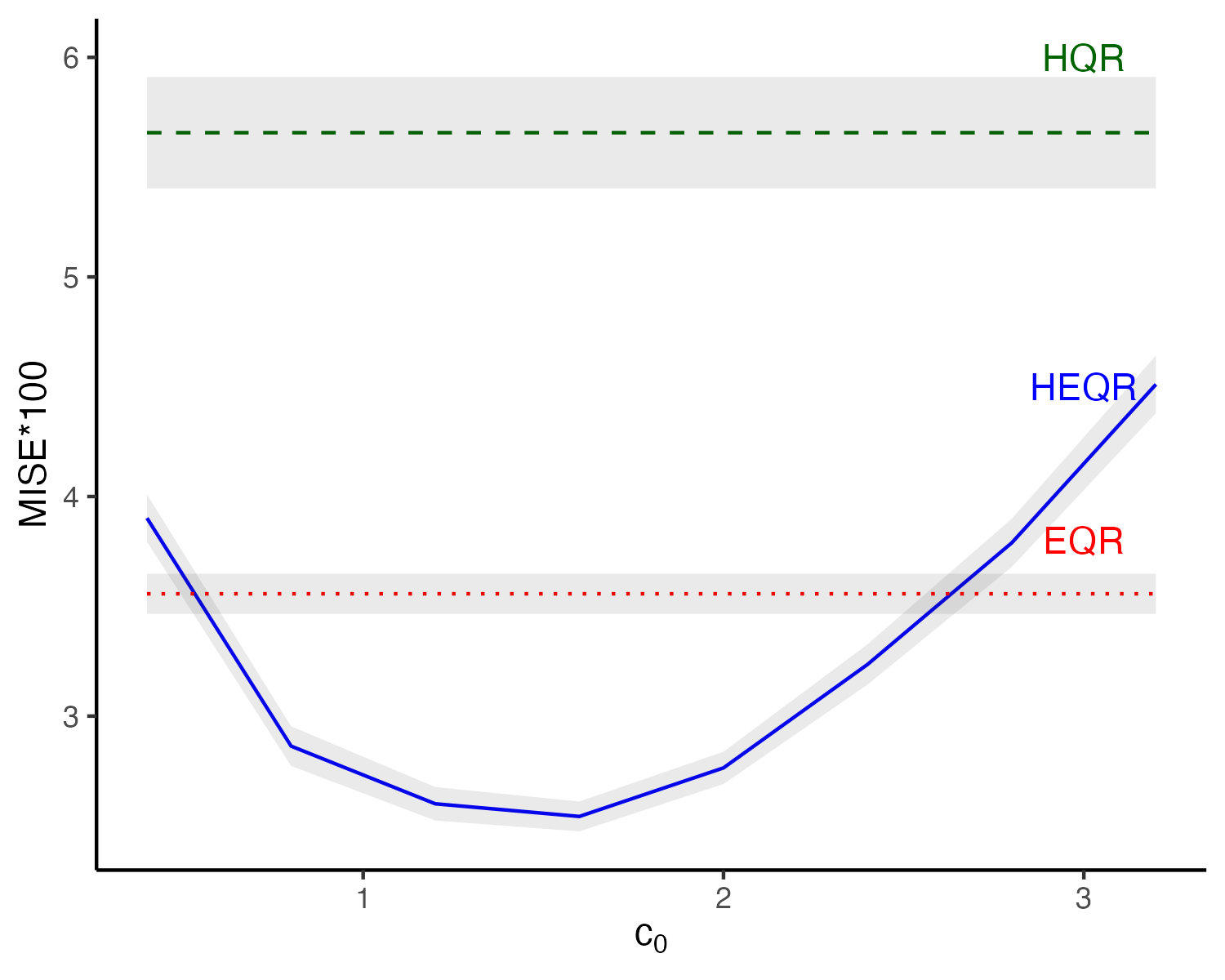}}
\subfigure[$n=5000$]{
	\label{Fig.c02}
	\includegraphics[width=0.45\textwidth]{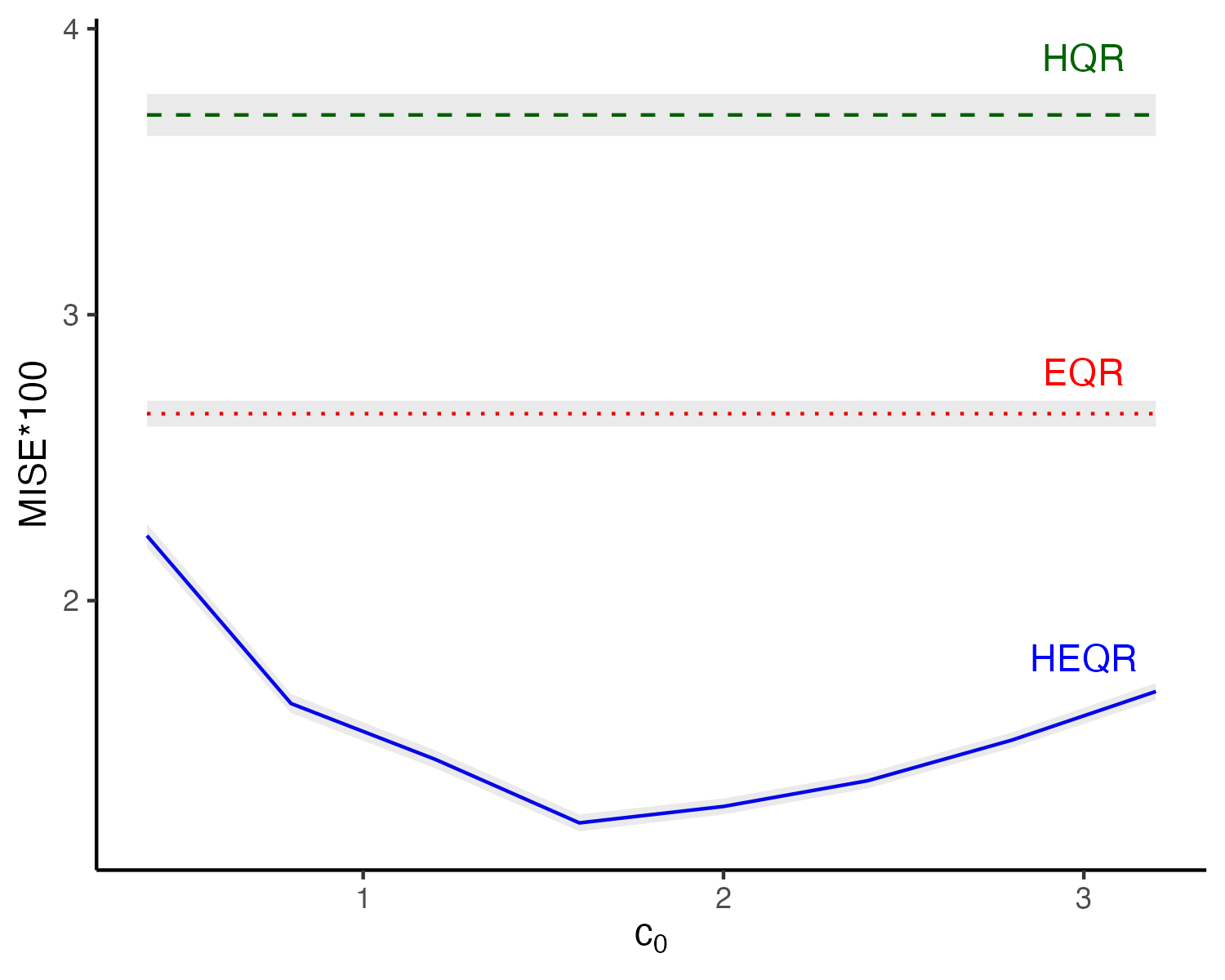}}
\caption{The MISE of estimators for the extreme conditional quantile at $\tau_n=0.995$ against $c_0$ in Case 1 with $n=1000$ (a) and $n=5000$ (b).}
\label{Fig:sensitivityc0}
\end{figure}

\begin{figure}
\centering  
\subfigure[$\tau_n=0.995$]{
	\label{Fig.MSEdelta1}
	\includegraphics[width=0.45\textwidth]{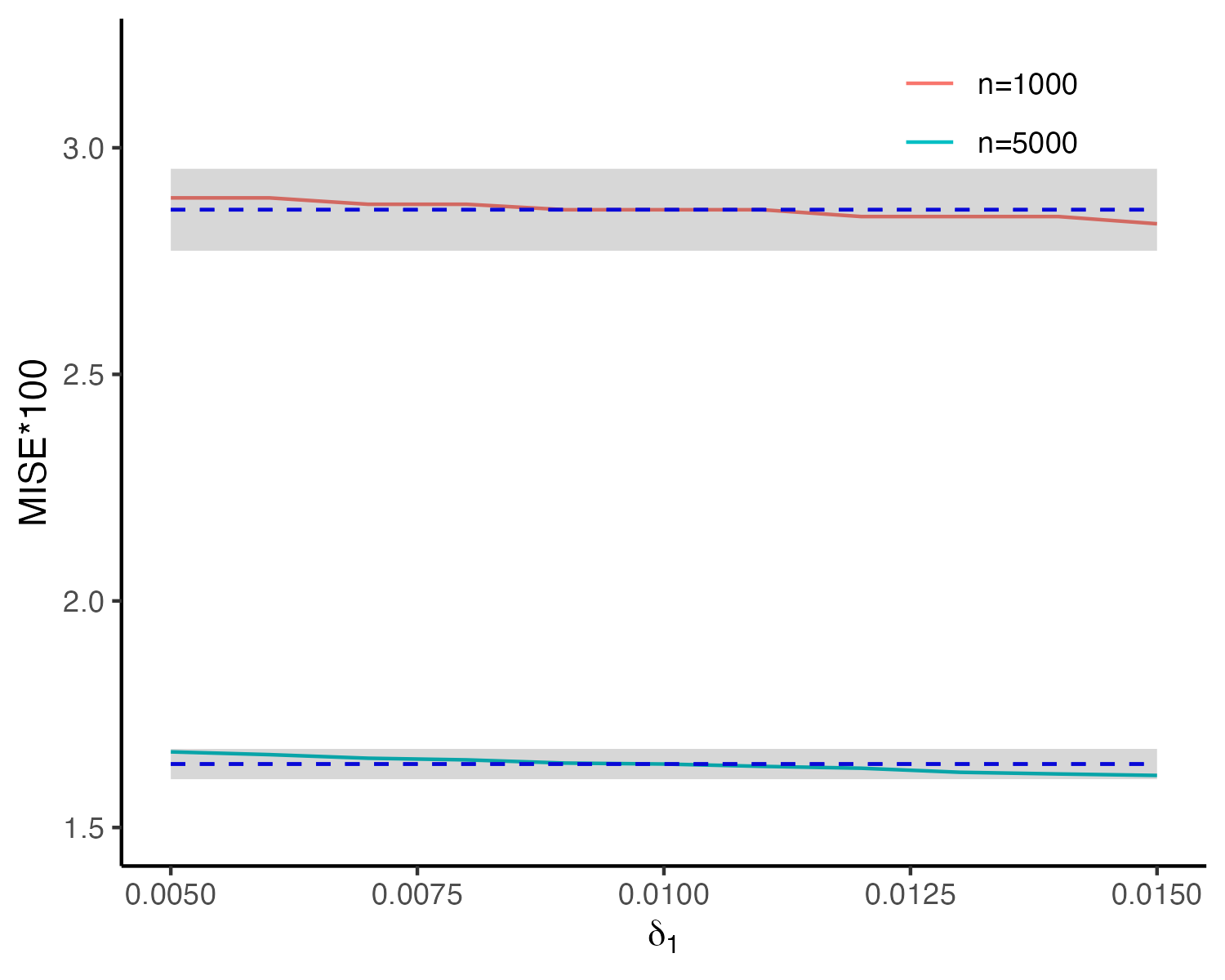}}
\subfigure[$\tau_n=0.999$]{
	\label{Fig.SDdelta1}
	\includegraphics[width=0.45\textwidth]{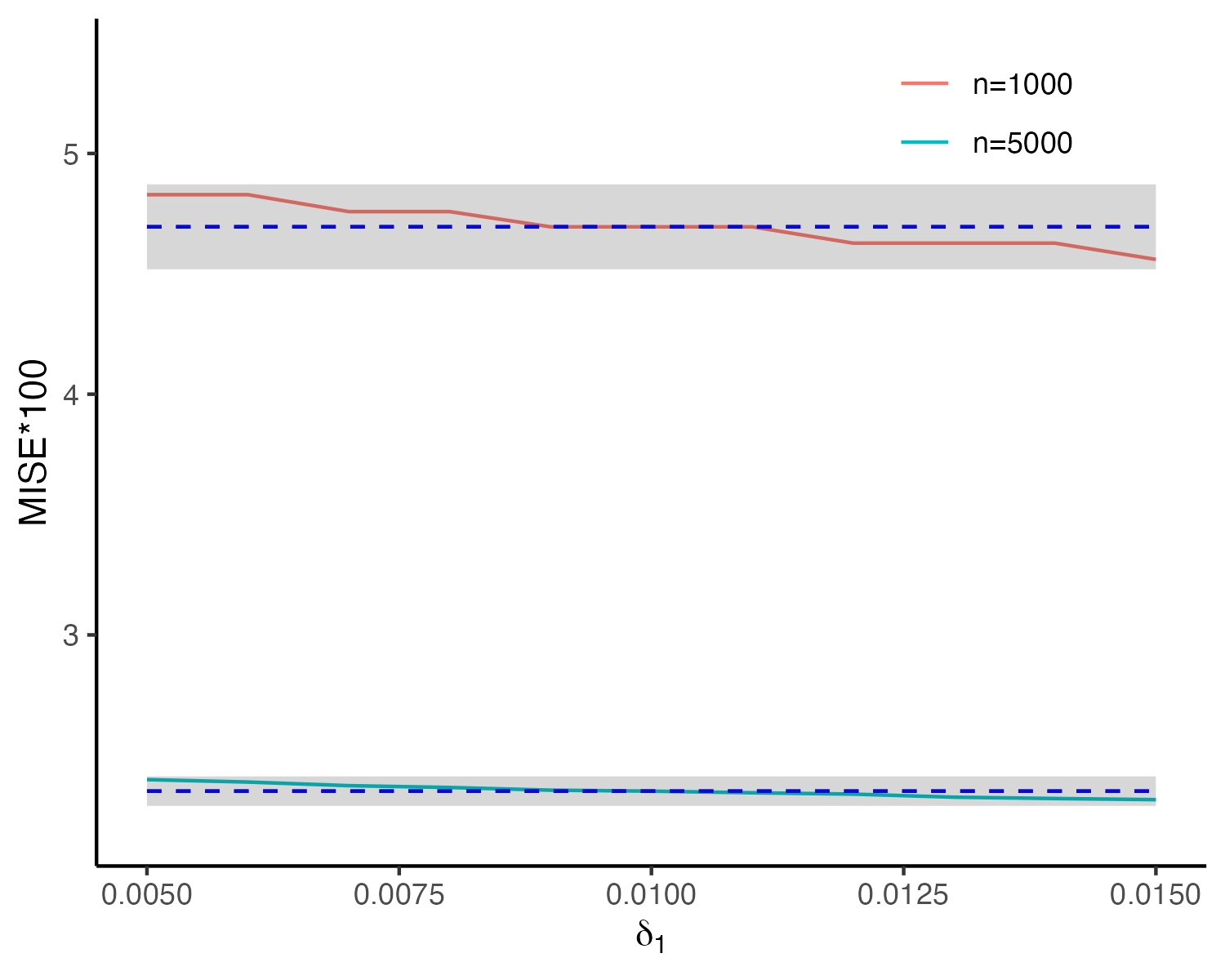}}
\caption{The MISE of HEQR estimator for the extreme conditional quantile at $\tau_n=0.995$ (a) and $0.999$ (b) against $\delta_1$ in Case 1.}
\label{Fig:sensitivitydelta1}
\end{figure}

\begin{figure}
\centering  
\subfigure[$\tau_n=0.995$]{
	\label{Fig.MSEdelta2}
	\includegraphics[width=0.45\textwidth]{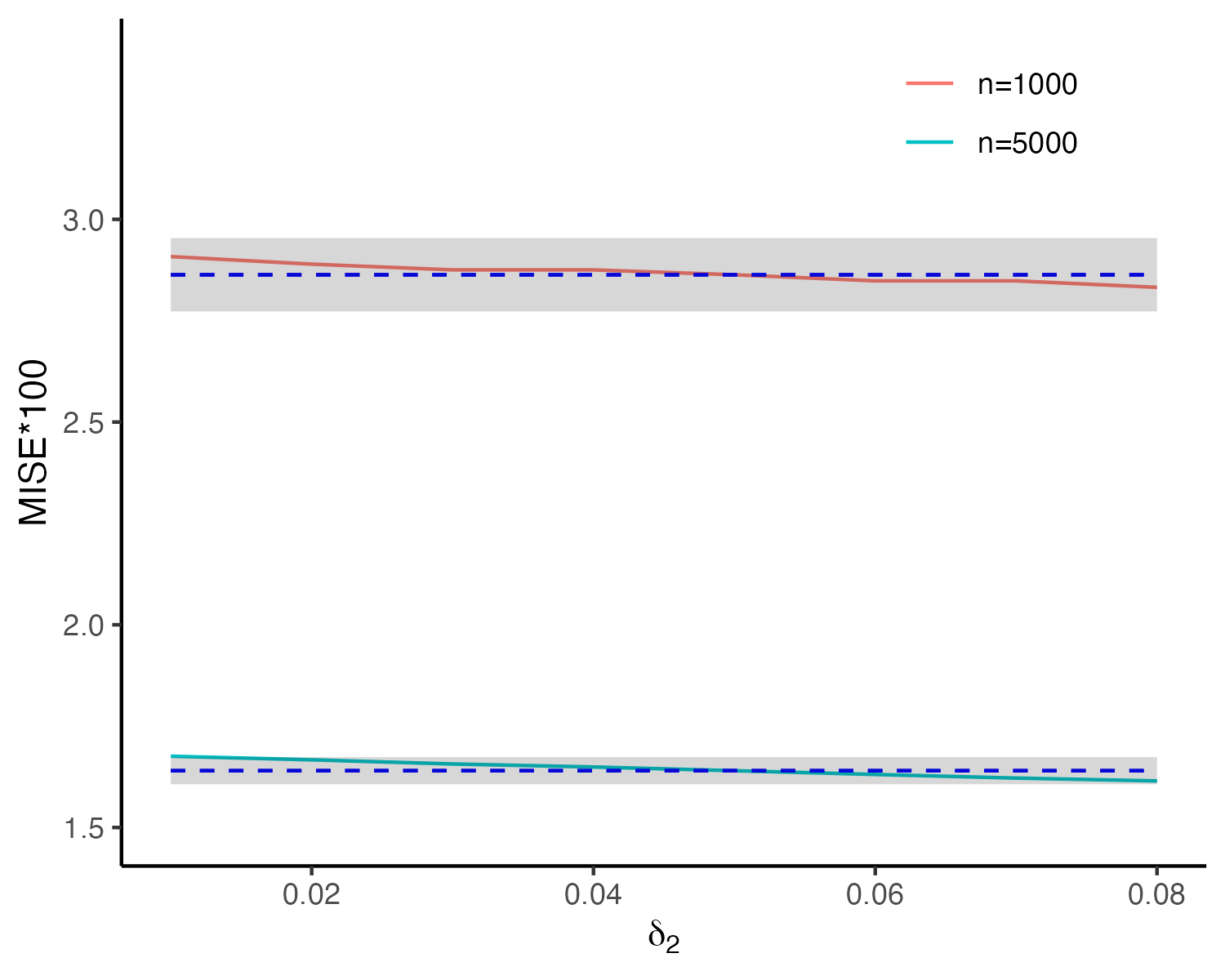}}
\subfigure[$\tau_n=0.999$]{
	\label{Fig.SDdelta2}
	\includegraphics[width=0.45\textwidth]{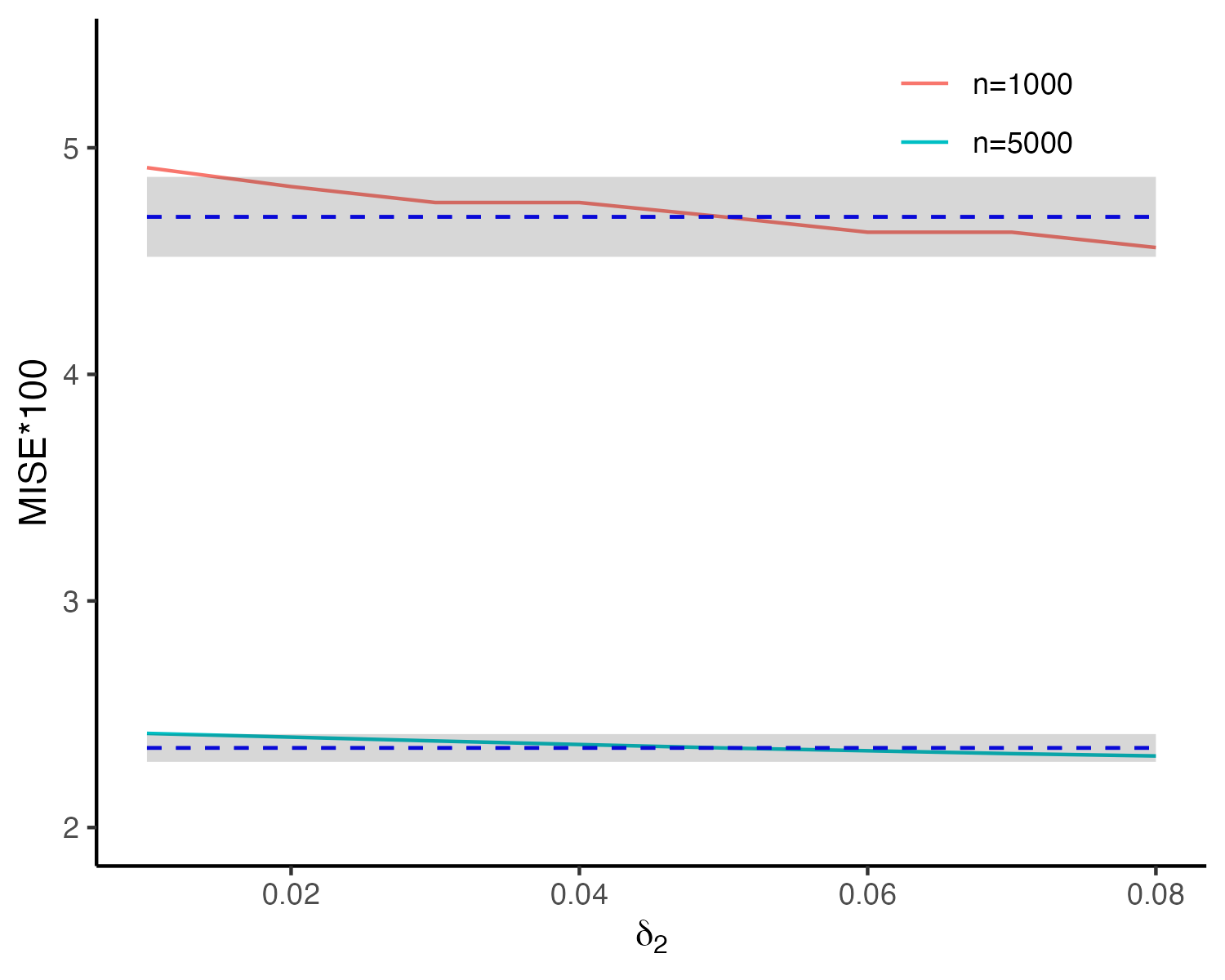}}
\caption{The MISE of HEQR estimator for the extreme conditional quantile at $\tau_n=0.995$ (a) and $0.999$ (b) against $\delta_2$ in Case 1.}
\label{Fig:sensitivitydelta2}
\end{figure}

\section{Analysis of Auto Insurance Claims}
\label{sec:application}
In this section, we analyze an auto insurance claims dataset to investigate the effects of various factors on the higher quantiles of claim amounts. The data is available on Kaggle at \url{https://www.kaggle.com/datasets/xiaomengsun/car-insurance-claim-data}. 
Insurance claims data are typically heavy-tailed and heterogeneous, reflecting the diverse nature of risks they embody and the array of factors that can influence claim sizes and frequencies. These factors not only complicate the analysis but also provide a rich vein of insights.}
{We consider the following quantile regression model:
$$Q_{Y_i}(\tau|\bX) = \beta_0(\tau) + X_{i1}\beta_1(\tau) + X_{i2}\beta_2(\tau) + \cdots + X_{ip}\beta_p(\tau), \quad i = 1, \ldots, n = 8423,$$
where \(Y\) represents the auto claim amount (in USD), and $X_{i1},X_{i2}, \ldots,X_{ip}$ are \(p = 43\) covariates included after preprocessing.
To account for nonlinearity, we include quadratic terms for ``Age" and ``Income," as well as quadratic and cubic terms for ``Vehicle Value" in the model, with both ``Income" and ``Vehicle Value"  on the log-transformed scale. A detailed description of the variables can be found in Section S.4 of the Supplementary Materials. We focus on high quantiles (\(\tau = 0.991, 0.995, 0.999\)) to study the impact of various factors on large claims.}


We conduct a cross-validation study to compare the performance of different methods for predicting the extremely high conditional quantiles of auto claims. We randomly divide the data set into a training set (20\% of the data set, $n_1=1684$ observations) and a testing set (the remaining 80\% of the data set, $n_2=6739$ observations). We apply HEQR, EQR, and HQR to analyze the training set and predict the extreme quantiles of auto claims conditional on the covariates in the testing set. {For HEQR, we set \( k = \lfloor 0.8 n_1^{0.51} (\log p)^{0.55} \rfloor \) and estimate \(\gamma\) at a central data point, with continuous covariates set to their mean values and binary variables set to their modes.} Since $\mathbbm{1}\{{Y}<Q_Y({\tau} | \mathbf{X})\}$ has a mean of $\tau$ and a variance of $\tau(1-\tau)$, this motivates us to consider the following prediction error (PE) metric: $$\mathrm{PE}=\{n_2 \tau(1-\tau)\}^{-1 / 2} \sum_{j\in\mathcal{I}}[\tau-\mathbbm{1}\{{Y}_j<\widehat{Q}_\tau({Y} | \mathbf{X}_j)\}],$$
where $\mathcal{I}$ is the index set of the testing set.
The cross-validation process is repeated 500 times. Table \ref{table:MedianAPE} summarizes the median absolute PE for different methods at $\tau_n = 0.991$, $0.995$, and $0.999$, with values in parentheses representing the corresponding median absolute deviation. The results indicate that HEQR yields significantly smaller prediction errors than the other two methods across all three quantile levels, demonstrating higher efficiency.


\begin{table}[!htb]
\begin{threeparttable}
\caption{The median absolute value (with median absolute deviation in parentheses) of prediction errors for different methods at $\tau_n = 0.991$, $0.995$, and $0.999$, based on cross-validation of the auto claims data.}
\renewcommand\arraystretch{1.5}
\label{table:MedianAPE}

\begin{tabular}{llll}
	\hline\hline
	& $\tau_n=0.991$ & $\tau_n=0.995$ & $\tau_n=0.999$ \\\hline
	HEQR & 1.84(1.13)     & 1.83(1.19)     & 1.28(0.63)      \\
	EQR   & 7.82(0.00)     & 5.82(0.00)     & 2.60(0.00)      \\
	HQR   & 8.95(2.06)     & 13.96(3.02)    & 32.86(6.74)     \\\hline\hline 
\end{tabular}
\begin{tablenotes}
	\tiny
	\item {HEQR}: the proposed estimator; {EQR}: the  method in \cite{wang2012estimation}; {HQR}: the high-dimensional qunatile regression estimator in \cite{belloni2011ℓ1}.
\end{tablenotes}
\end{threeparttable}
\end{table}

\begin{figure}
\centering  
\includegraphics[width=0.7\textwidth]{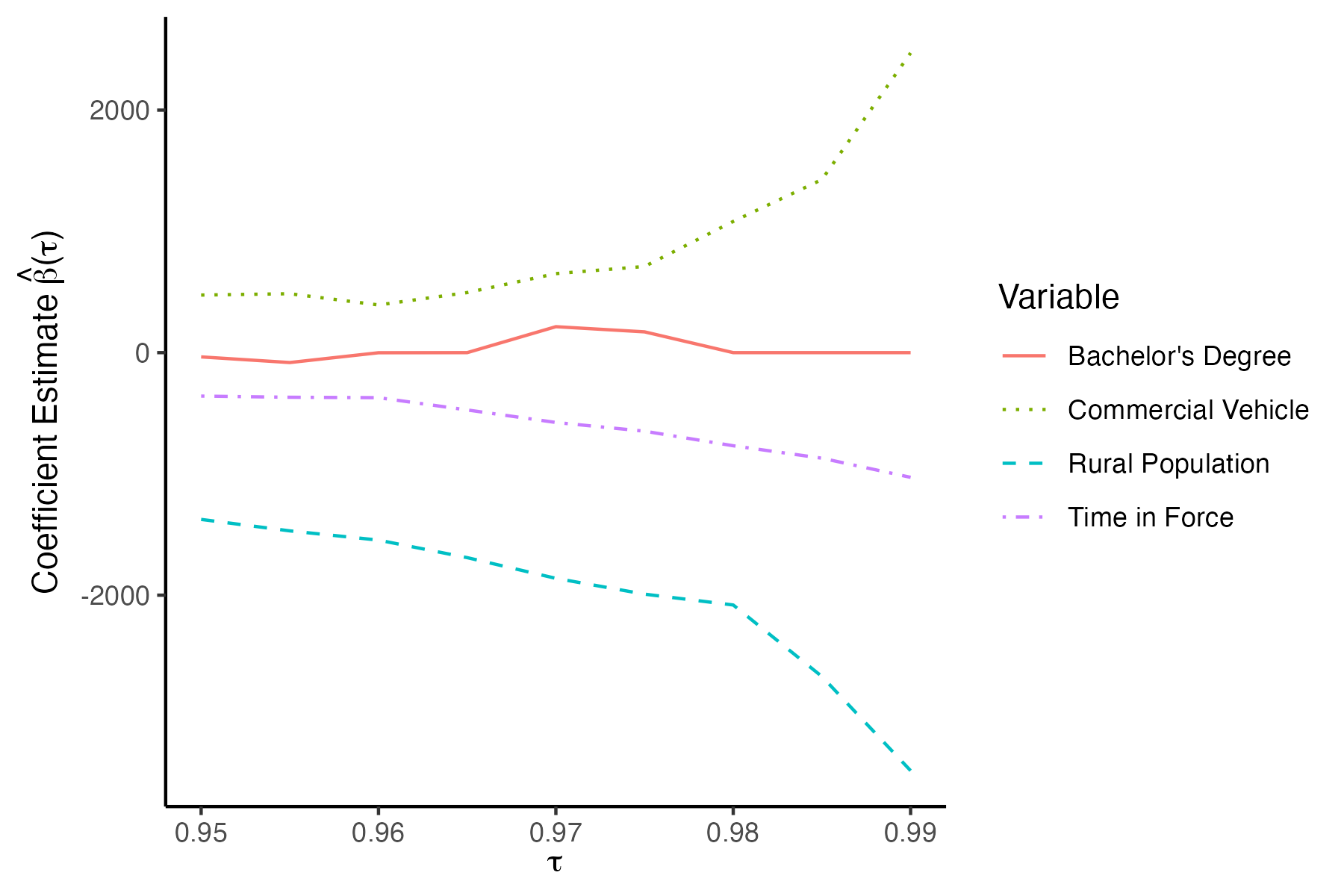}
\caption{The estimated quantile coefficients of selected covariates for the auto claims data versus the quantile level at the upper tail.}
\label{Fig:coeff}
\end{figure}

Next, we analyze the full dataset to estimate extreme quantiles and identify important variables using HEQR. {
Out of the 43 covariates, our method identified 34 to have nonzero effects at the intermediate high quantile level at $\tau_{0n}=1-k/n$ with $k = \lfloor 0.8 n^{0.51} (\log p)^{0.55} \rfloor$ and $n=8423$. Notably, one of the most significant variables identified by the HEQR model is whether an individual resides in a rural district. This finding aligns with prior research by \cite{clemente2023modelling}, which links rural residences to lower auto insurance claims due to reduced claim frequency and severity. Additionally, our results show that an increase in vehicle age corresponds to a decrease in claims, consistent with \cite{clemente2023modelling}, who also found vehicle age negatively impacts claim frequency.} 
Figure \ref{Fig:coeff} shows the $\ell_1$-penalized quantile coefficient estimates $\widehat\bbeta_j(\tau)$ for selected covariates at upper quantiles. The results indicate that being a commercial vehicle positively affects auto claims, with the effect increasing at higher quantiles. In contrast, ``Rural Population" and ``Time in Force" (the duration the insurance policy has been in effect, abbreviated as TIF) have negative effects, which become stronger at higher quantiles. The variable ``Bachelor's Degree" shows little impact on auto claims. These findings highlight the heterogeneity of covariate effects, particularly in the upper tail of the claim distribution.



\begin{figure}
\centering  
\subfigure[$\tau_n=0.995$]{
\label{Fig.sub.11}
\includegraphics[width=0.45\textwidth]{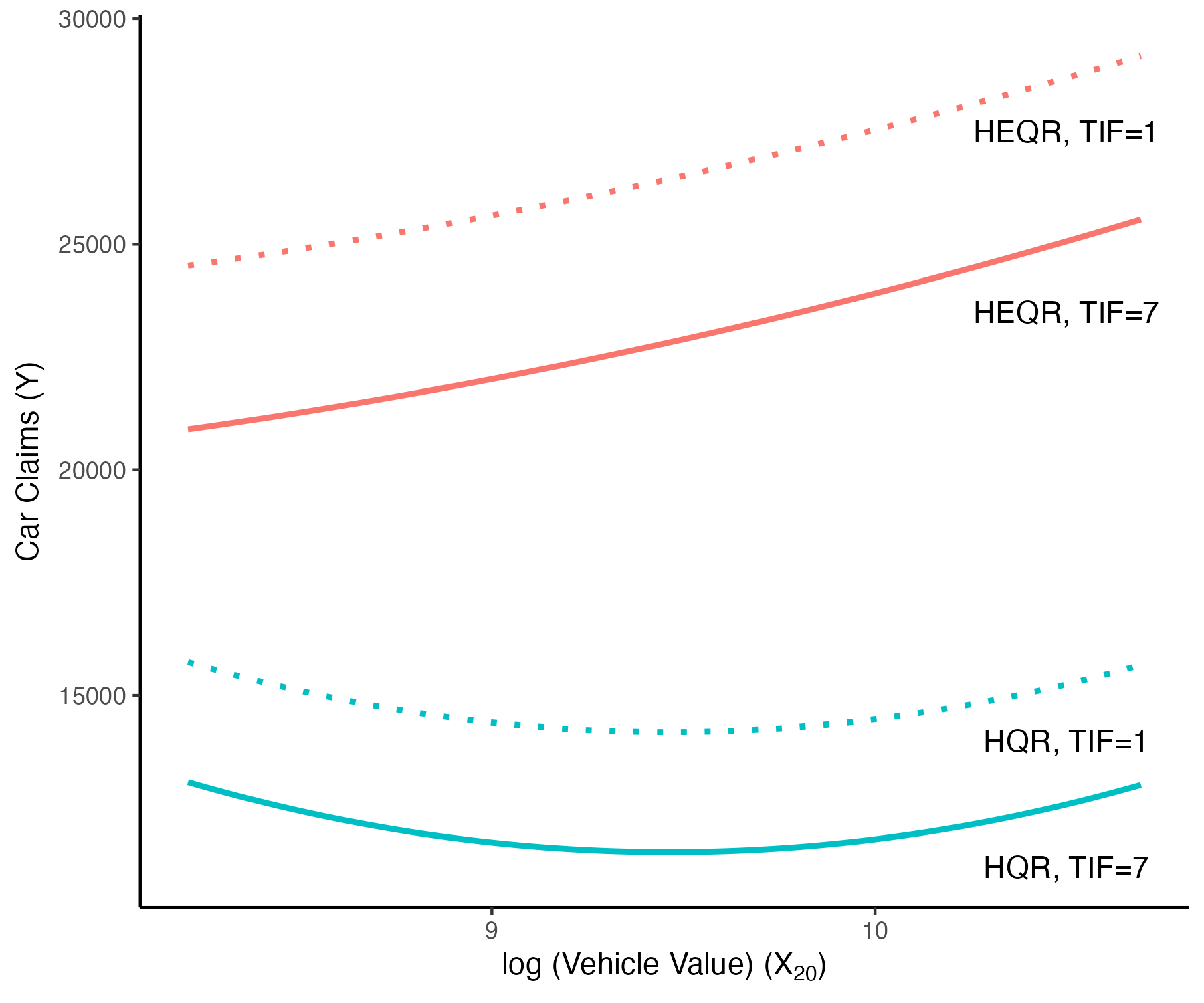}}
\subfigure[$\tau_n=0.999$]{
\label{Fig.sub.13}
\includegraphics[width=0.45\textwidth]{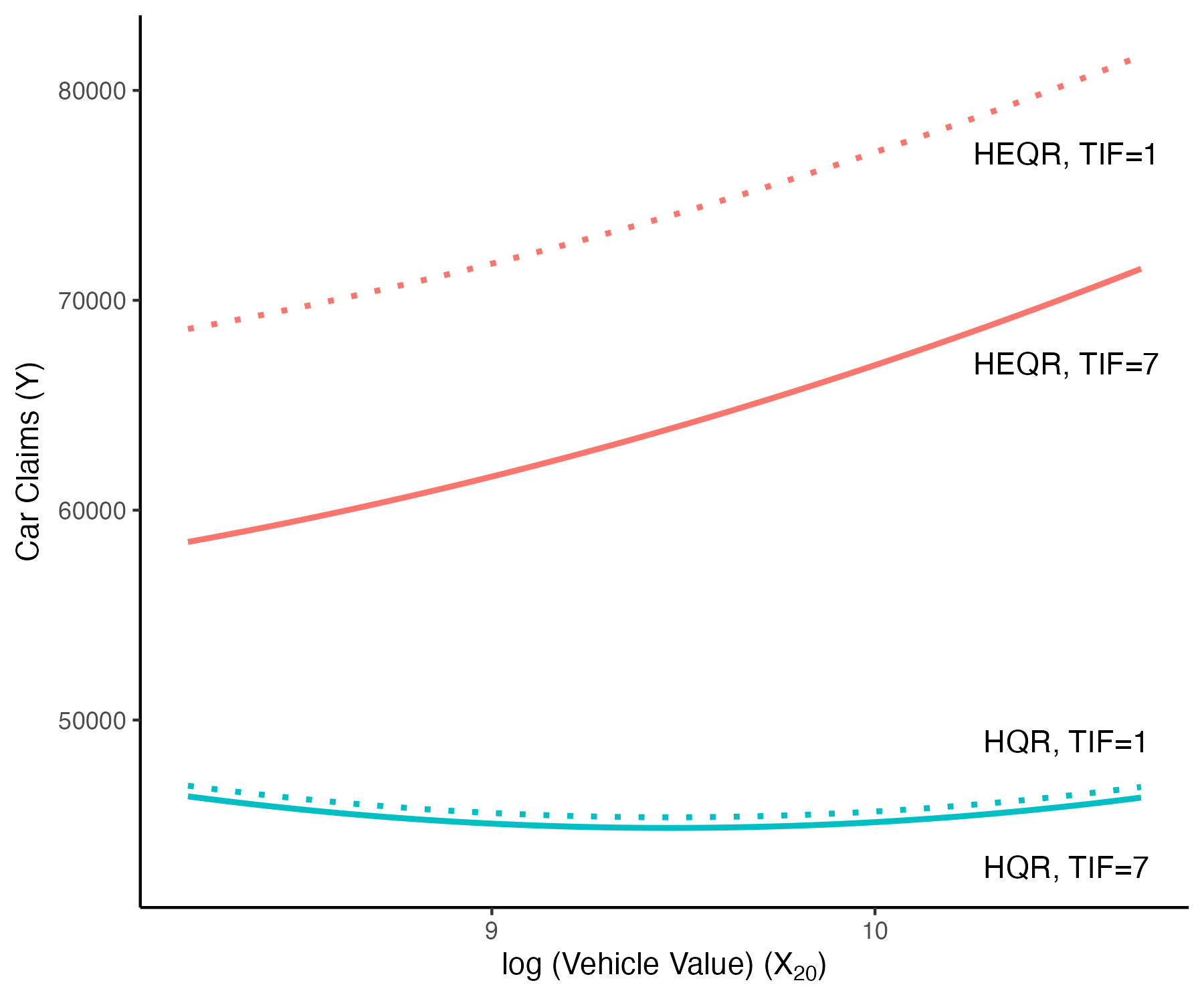}}
\caption{The estimated extreme conditional quantiles of auto claims (in \$) from HEQR and HQR against the log-transformed Vehicle Value ($X_{20}$) for two groups: time in force equal to 1 (TIF=1) and 7 (TIF=7), at $\tau_n=0.991$ and $0.999$, with other continuous covariates set to their mean values and categorical variables to their modes.}
\label{Fig:SP}
\end{figure}

{
We also include the full-data analysis results from HQR for comparison. Figure \ref{Fig:SP} presents the extreme quantile estimates of auto claims at $\tau_n = 0.991$ and 0.999 as a function of $X_{20}$, log-transformed Vehicle Value, for two groups of individuals with different TIF from both HQR and HEQR, with other continuous covariates set to their mean values and categorical variables to their modes. The two groups correspond to individuals with TIF of 1 year and 7 years, representing the 0.25 and 0.75 quantiles, respectively. Results from HEQR suggest that, for both groups, higher vehicle values lead to larger claims, likely due to higher repair costs for more expensive vehicles, which aligns with common expectations and prior findings \citep{clemente2023modelling}. In contrast, HQR shows little variation in high quantiles of claims across vehicle values. Both methods indicate that individuals with a shorter TIF tend to have higher claims at $\tau_n = 0.991$. This finding aligns with our common understanding, as shorter TIFs may reflect less experienced or stable policyholders, potentially leading to higher-risk profiles and more frequent or severe claims. HEQR consistently shows the effect of TIF across high quantiles, while HQR becomes unstable at $\tau_n = 0.999$, displaying almost no difference between the two groups.}

\bigskip
\begin{center}
{\large\bf SUPPLEMENTARY MATERIAL}
\end{center}

\begin{description}

\item[Supplementary Material for High-dimensional Extreme Quantile Regression:] The supplementary material includes all the technical conditions, proofs of theorems, additional simulation results, and further details on the analysis of the auto insurance claims data. (.pdf file)

\item[Auto Insurance Claims Data Set:] Data set used in the illustration of the proposed methods in Section \ref{sec:application}. (.csv file)
\end{description}


\bibliographystyle{agsm1}
\bibliography{Bibliography-MM-MC}
\end{document}